# Characterization of high-quality MgB$_2$(0001) epitaxial films on Mg(0001)

Luca Petaccia[1], Cinzia Cepek[2], Silvano Lizzit[1], Rosanna Larciprete[1,3], Roberto Macovez[2], Massimo Sancrotti[2,4] and Andrea Goldoni[1,*]

[1]Sincrotrone Trieste S.C.p.A., s.s. 14 km 163,5 in Area Science Park, 34012 – Trieste, Italy

[2]Laboratorio TASC-INFM, s.s. 14 km 163,5 in Area Science Park, 34012 – Trieste, Italy

[3]CNR-Istituto dei Sistemi Complessi, Via Salaria Km 29.3, 00016 Monterotondo (RM), Italy

[4]Dipartimento di Matematica e Fisica, Universita` Cattolica del Sacro Cuore, Via dei Musei 41, 25121 – Brescia, Italy

**ABSTRACT**

High-grade MgB$_2$(0001) films were grown on Mg(0001) by means of ultra-high-vacuum molecular beam epitaxy. Low energy electron diffraction and x-ray diffraction data indicate that thick films are formed by epitaxially oriented grains with MgB$_2$ bulk structure. The quality of the films allowed angle-resolved photoemission and polarization dependent x-ray absorption measurements. For the first time, we report the band mapping along the Γ-A direction and the estimation of the electron-phonon coupling constant λ=0.55±0.06 for the surface state electrons.

**1. Introduction**

The unexpected discovery of superconductivity in $MgB_2$ at $T_C$ =39 K [1], almost twice the temperature of other simple inter-metallic compounds, has sparked an endeavor to uncover its basic physical properties, such as the mechanism of the superconductivity, and the various aspects of its synthesis finalized to the application in superconductor-based devices.

Although $MgB_2$ powders can be easily obtained from low cost and non-toxic reagents, the synthesis of high-quality (phase-pure) samples is still very challenging. Different synthetic techniques and different thermodynamic parameters produce polycrystalline samples with slightly different electronic and superconducting characteristics, mainly ascribed to the effect of impurities, structural defects (i.e. Mg vacancies) and lattice strains [2-7]. These facts, together with the rapid oxidation of $MgB_2$ in ambient atmosphere, are considered the main reason for the discrepancies between the experimental data reported for polycrystalline samples.

The growth of high-grade single crystals [8] has improved the quality of the experimental data and opened a route for various important physical studies [8-17]. However, the sub-millimeter size of these crystals and the need of exposing the samples to air before any experimental investigation, have limited the application of many techniques as, for example, angle-resolved photoemission spectroscopy (ARPES), which is the most general and uniquely powerful tool for the direct investigation of the occupied band structure of solids. Photoemission experiments, as well as other basic investigations like scanning tunneling microscopy and spectroscopy or transport measurements, will take great advantage by the development of an adequate ultra-high-vacuum (UHV) *in situ* growth of ordered phase-pure $MgB_2$ thin films.

Here we show that this can be achieved by UHV molecular beam epitaxy (MBE).

Ordered thin films of MgB$_2$ were epitaxially grown on a cm$^2$-sized Mg(0001) substrate by co-deposition of Mg and B in UHV [18]. Epitaxy is possible owing for the small mismatch (about 3.5%) between the in-plane lattice parameters of Mg(0001) (|a|=|b|=3.191 Å) and MgB$_2$ (|a|=|b|=3.085 Å). Co-deposition of Mg and B (atomic flux ratio ~ 2:3) on the clean substrate held at 493 K allows the layer-by-layer formation of ordered MgB$_2$ films as confirmed by low-energy electron diffraction (LEED), x-ray diffraction, x-ray photoemission (XPS) and absorption (XAS) spectroscopies, and ARPES.

**2. Experimental**

The photoemission and x-ray absorption experiments were performed in the UHV end-station of the SuperESCA beamline at Elettra (base pressure 5x10$^{-11}$ mbar). The photoelectrons were collected using a double-pass hemispherical electron energy analyzer with an angular resolution of ±0.5˚. The angle between the incident light and the analyzer is fixed to 70˚. The overall energy resolution was about 40 meV for valence band measurements and better than 200 meV for core level spectra.

The x-ray diffraction and surface x-ray diffraction data were collected in the UHV end-station of the ALOISA beamline at Elettra at fixed scattering geometry by measuring the intensity of the (002) and (110) reflections, respectively, as a function of the photon energy.

The Mg(0001) substrate was cleaned and ordered by subsequent cycles of sputtering and annealing at 493 K. Pure metal sources of Mg slugs (99.95%) and B wires (99.5%) were used. Mg was evaporated from a resistively heated Ta cell and B was evaporated using an electron-beam gun. Mg and B evaporators were carefully cleaned with several days of degassing. During the co-deposition of Mg and B the pressure was maintained in the range 1x10$^{-9}$ mbar < p< 2x10$^{-9}$ mbar. The evaporation rate was such as one complete layer of MgB$_2$ (one plane of Mg + one plane of B$_2$) is formed in 7 min. The evaporation rates were determined on the basis of the attenuation of the substrate peaks and the growing intensity of the evaporated chemical elements as a function of the evaporation time in XPS by depositing B on clean Mg(0001) and Mg on a clean copper plate at room temperature. As shown elsewhere [18], the film growth proceeds layer-by-layer.

**3. Results and discussion**

Figure 1 shows the XPS spectrum, at 655 eV of photon energy, of a film obtained after 2h of co-evaporation. The estimated film thickness is equivalent to 18 $MgB_2$ layers [18]. Only B and Mg peaks are visible confirming that the presence of contaminants, like carbon or oxygen, is below the detection limit of XPS (0.2%). The area ratio (Mg 2*s* / B 1*s*) = 0.30±0.02 is consistent with a stoichiometric $MgB_2$ film, mainly Mg terminated, and suggests that the Mg signal only comes from Mg reacted with B, while the signal from the Mg(0001) substrate is completely hidden by the thickness of the film. This is supported by the change in line shape occurring to the Mg core level peaks after the film deposition. The inset of Fig. 1a compares the Mg 2*s* photoemission peak before (clean Mg) and after film deposition. Binding energy and lorentzian width remain practically unchanged (88.64±0.04 eV and 0.50±0.03 eV, respectively) after deposition, while the Mg 2*s* core level of the co-deposited film has larger (doubled) Doniach-Sunjic asymmetry and Gaussian width compared to clean Mg. This behavior is in agreement with an increased density of states at $E_F$ and the excitation of softer phonons, as expected on passing from Mg to $MgB_2$.

The inset of Fig.2 shows the LEED pattern observed at 69 eV of electron beam energy for a film obtained after 2h of co-deposition. At this kinetic energy, the electron probing depth is less than the photoelectron escape-depth in the XPS spectra shown in Fig.1, therefore, the substrate is not probed by LEED. The pattern exhibits hexagonal symmetry and, within our incertitude, the corresponding surface lattice parameters are consistent with the in-plane parameters of $MgB_2$ (3.085 Å). Because of the coherence length of the low energy electrons and the fact that we observe a single domain exagonal LEED pattern, our film must be formed by oriented epitaxial grains at least few hundreds Å of lateral dimension.

A more precise determination of the lattice parameters of our films can be obtained by XRD measurements. Fig. 3a shows the results about the in-plane lattice parameter obtained by collecting the x-ray intensity of the (110) reflection as a function of the photon energy for a film thickness equivalent to 18 $MgB_2$ layers. The peak at lower photon energy corresponds to the (110) reflection of the Mg(0001) substrate, while the less intense peak at higher photon energy is due to the (110) reflection of the epitaxial film. The position of this second peak, with respect to the substrate reflection, corresponds to a lattice parameter of 3.07±0.02 Å, compatible with $MgB_2$ bulk lattice parameter a = 3.085 Å. Although the crystalline quality of the substrate is not very good (as suggested by the large width of its in-plane reflection), from the width of the (110) reflection of the epitaxial layer we can assess that the film is formed by grains whose lateral dimension is at least 150 Å, compatible with the fact that we observe a LEED pattern.

The perpendicular c-axis parameter has been obtained by measuring the (002) reflectivity as a function of the photon energy, as shown in Fig. 3b. From the position of the epitaxial film reflection peak with respect to the substrate Mg(0001) peak, we obtain a perpendicular lattice parameter of 3.48±0.07 Å, in agreement with the $MgB_2$ bulk lattice parameter c = 3.52 Å. Moreover, a more detailed analysis of the diffraction data as a function of the film thickness (not shown here) [19] demonstrates that, while at the beginning the c-axis parameter is contracted and the in-plane parameter is expanded to match the Mg substrate parameters, the crystal structure becomes that of bulk magnesium diboride after the growth of about 15 layers. Indeed, the full set of data discussed above suggests that the co-evaporated epitaxial film equivalent to 18 MgB$_2$ layers has a bulk MgB$_2$ structure.

A further indication that the film is actually MgB$_2$ comes from the x-ray absorption spectra at the B 1$s$ threshold shown in Fig.2. These spectra were measured collecting the boron Auger electrons at 180±4 eV of kinetic energy and represent, therefore, the B-projected empty states. The features present in the XAS spectra compare well with calculations including the presence of the core-hole [20, 21] and with the B 1$s$ absorption spectra measured with the electron beam of transmission electron microscopes on single crystalline MgB$_2$ grains [20, 22]. Our film shows a strong polarization dependence of the XAS spectra, in good agreement with theoretical calculations for the (0001) surface of MgB$_2$ single crystals (bottom curves) [21], the first absorption peak disappearing when the linear electric field of the light polarization is parallel to the surface plane (i.e. perpendicular to the c-axis). Similar spectra and polarization behavior were also observed by acquiring XAS spectra in total yield mode, which is more bulk sensitive, pointing to comparable quality of the layers underneath the surface.

The high-grade of our MgB$_2$ films is confirmed also by its occupied band structure measured using ARPES. Band mapping based on direct transitions in angle-resolved photoemission experiments is a direct probe of the electronic structure and of the electron scattering processes, providing both volume- and surface-sensitive information of fundamental importance to completely describe the electronic properties of solids. Extensive studies performed during the last three years demonstrate that, despite the chemical and structural simplicity of MgB$_2$, the band structure is extremely sensitive to the MgB$_X$ phase formed [23], to the length [24] and distortion [25] of the lattice constants, and to the presence of substitutional impurities [26, 27]. Therefore, it can be used as a fingerprint of pure MgB$_2$ formation.

The occupied bands of MgB$_2$ consist of bonding σ-bands made from in-plane sp$^2$ hybrids in the boron layer, bonding π and anti-bonding π* bands [23-28] formed by boron $p_z$ orbitals. Compared to the iso-structural graphite, the π-band is pushed to lower binding energy and displays a marked three-dimensional character due to the reduced c/a axis ratio. The important consequence is a charge transfer from the σ to the π-bands which creates holes at the top of the bonding σ-bands. There are, therefore, two kinds of bands crossing the Fermi level producing two superconducting gaps with different characteristics [13, 15, 16, 23, 28]: σ-bands show a superconducting gap larger than π-bands. Interestingly, on good single crystal samples a surface state band, displaying a superconducting gap of size comparable to that measured for the σ-bands, has been observed [15].

Fig. 4 shows the band dispersion of our film along the Γ-A direction ($k_\perp$), obtained by changing the photon energy from 95 eV to 185 eV (step 5 eV) and collecting the photoelectrons at normal emission ($k_{//}$ = 0). There is a couple of dispersing bands at binding energies higher than 2.5 eV and non-dispersing features at ~ 1.6 eV and 3.2 eV that correspond to surface states. The couple of dispersing bands is due to bulk π- and σ-states showing opposite dispersions [23]. The intense surface state at ~ 1.6 eV of binding energy matches the surface band calculations for the most thermodynamically stable Mg terminated surface of $MgB_2$(0001) at $k_{//}$=0 [29, 30]. The remaining non-dispersing peak at ~ 3.2 eV is almost degenerate with the bulk π-band at Γ and it corresponds to the surface state for the B-terminated surface, indicating the presence of minority $MgB_2$ domains having this termination. There is indeed a third non-dispersing feature at about 6 eV, due to O $2p$ states, indicative of a small oxygen contamination growing with time. The presence of this feature, however, does not affect our main observations and conclusions.

The calculated bulk band structure of $MgB_2$ [23] is superimposed as circles to the two-dimensional intensity plot of our data in Fig. 4b. The very good agreement with calculations and the fact that tentative measurements of the $k_\perp$ band structure previously failed on MgB$_2$ single crystals [9, 15] confirm the high-quality bulk structure of our film.

In Fig. 5 we show the in-plane band dispersion measured at hν=105 eV along the "Γ-K-M-K-Γ" direction at k$_\perp$ ~ 0.10 Å$^{-1}$, obtained by rotating the polar angle of the sample every 3˚. Once again, we compare our experimental results with theoretical calculations for both bulk [23] and surface [29, 30] bands of MgB$_2$. Although the experimental apparatus where we performed the measurements is not optimized for valence band angle-resolved photoemission experiments, the agreement with the calculated MgB$_2$ band structure is good over a wide binding energy and momentum range, never tested before.

The high surface quality of our MgB$_2$ films is supported by the very intense and sharp surface state near the Fermi level. This surface state has π-character and originates mainly from the boron layer underneath the topmost Mg surface layer [30]. The fact that at normal emission and using a photon energy of 105 eV this surface peak is very intense and well separated from any other feature in the photoemission spectra allows investigating the temperature dependence of its linewidth. Fig. 6a shows the photoemission spectra measured at normal emission (Γ point) as a function of temperature. Fig. 6b reports the temperature dependent width of the surface state peak, as obtained by fitting the spectra to a bulk integrated density of states, plus a lorentzian peak (surface state) and a gaussian peak (tail of the π-band), all multiplied by the Fermi-distribution function and convoluted with a gaussian function which simulates our temperature independent experimental energy resolution (70 meV).

Within the quasi-particle picture of the electron-phonon coupling (valid very close to or very far from E$_F$ on the phonon bandwidth energy scale) and the Debye's model [31], the temperature dependent width W$_s$ of the surface state at ~ 1.6 eV of binding energy can be modeled as:

$$W_S(T) = 2\pi k_B \int_0^{\theta_D} \frac{\lambda}{2} \frac{\theta}{\theta_D}[1 + \frac{2}{\exp\left(\frac{\theta}{T}\right)-1}]d\theta \qquad (1)$$

where $k_B$ is the Boltzman constant, $\theta_D$ the Debye temperature, $\lambda$ the dimensionless electron-phonon coupling constant, and the Eliashberg function has been assumed proportional to the phonon density of states in two-dimensions, i.e. $\alpha^2F(\theta)=\lambda\theta/2\theta_D$. By fitting the data of Fig. 4b to equation $(1)$, assuming $\theta_D$ in the range 750 – 1200 K [25, 26] we obtain $\lambda=0.55\pm0.06$. This value is larger than the isotropic electron-phonon coupling reported for the electrons in the $\pi$-bands $\lambda_\pi = (\lambda_{\pi\pi} +\lambda_{\pi\sigma})N_\pi/N \sim 0.28\pm0.07$, while is comparable to the isotropic electron-phonon coupling for the $\sigma$-electrons $\lambda_\sigma = (\lambda_{\sigma\sigma}+\lambda_{\sigma\pi})N_\sigma/N \sim 0.50\pm0.11$, where $N=(N_\pi+N_\sigma)$ is the total density of states and $N_\pi/N_\sigma=1.37$ [13, 23, 25, 28]. Although the surface band mainly originates from boron $\pi$-states [30], the estimated $\lambda$ is in agreement with an increased density of states at the surface and with the photoemission observation of a bigger superconducting gap for the surface state electrons than for $\pi$-bands [15], comparable in size with the gap measured for the $\sigma$-electrons.

## 4. Conclusions

Although we have not measured superconducting properties of our films, we have demonstrated using LEED, x-ray diffraction, the polarization dependence of XAS and ARPES that the low temperature co-deposition of B and Mg in UHV allows the formation of ordered $MgB_2$ films on Mg(0001), with crystal structure like that of $MgB_2$ bulk samples. Our studies testify that the $MgB_2$ film, probably formed by oriented grains of about 150 Å of lateral dimension, grows with the c-axis perpendicular to the Mg(0001) surface plane. Thanks to the high surface quality, we were able to measure the in-plane band dispersion and to estimate the electron-phonon coupling constant for the surface band electrons. The band dispersion along the Γ-A direction (perpendicular to the surface plane), in good agreement with calculations, was also measured for the first time, further testifying the crystal periodicity also along the c-axis.

**Acknowledgements**

We would like to express our gratitude to the ALOISA staff for their help during the x-ray diffraction measurements and for several useful discussions. We are also grateful to G. Profeta for critical comments on this subject.

FIGURE CAPTIONS

**Figure 1:** XPS spectrum of the film grown after two hours of Mg and B co-deposition on Mg(0001). The inset compares the Mg 2$s$ spectrum of the film with that of the clean Mg. The corresponding fits, obtained using a Doniach-Sunjic (DS) lineshape convoluted with a Gaussian function, are also shown.

**Figure 2:** XAS spectra at the B 1s threshold measured as a function of the angle (indicated for each spectrum) between the linear light polarization and the normal to the surface (c-axis), for the film grown after two hours of Mg and B co-deposition on Mg(0001). The XAS spectra are compared to the electron energy loss spectrum across the B1s threshold (top spectrum) as measured on a $MgB_2$ single crystalline grain [22] and to the calculated $MgB_2$ polarized B 1s absorption spectra (bottom curves) for polarization parallel (red line) and perpendicular (black line) to the c-axis [21]. The inset shows the corresponding LEED pattern obtained with primary electron beam energy of 69 eV.

**Figure 3:** Grazing incidence x-ray diffraction data obtained by collecting the (110) reflection **(a)** and the (002) reflection **(b)** as a function of the photon energy. The positions of the Mg(0001) substrate peaks (corresponding to an in-plane parameter of 3.19 Å and a perpendicular parameter c/2=2.59 Å) and of the epitaxial overlayer equivalent to 18 $MgB_2$ layers (corresponding to an in-plane parameter of 3.07±0.02 Å and a perpendicular parameter c =3.48±0.07 Å) are marked.

**Figure 4:** Band dispersion along the Γ-A direction. The Brillouin zone is also shown. **(a)** Valence band spectra collected at normal emission ($k_{//}$=0) as a function of the photon energy (from 95 eV up to 185 eV every 5 eV) of the 18 ML film co-deposited at 493 K. The surface states are indicated. **(b)** Calculated $MgB_2$ band structure (dots)

along the Γ-A direction [23] superimposed to the bi-dimensional plot of our experimental data. Black corresponds to the highest intensity.

**Figure 5:** Band dispersion along the Γ-K-M-K-Γ direction at hν=105 eV. **(a)** Valence band spectra of the 18 ML co-deposited film collected as a function of the polar angle (every 3˚).

**(b)** Calculated projected MgB$_2$ bulk states [23, 29, 30] - red and white dots - and Mg-terminated surface [29, 30] band – yellow dots - along the Γ-K-M direction superimposed to the bi-dimensional plot of our experimental data. Black corresponds to the highest intensity.

**Figure 6: (a)** Normal emission valence band photoemission spectra obtained at hν=105 eV as a function of temperature. As an example, the fit of the spectrum at 403K (red curve) together with the fitting components (blue = surface state, pink = π-band, green = bulk integrated DOS) is also shown. **(b)** Surface state linewidth as a function of temperature. The corresponding fit using equation *(1)* and θ$_D$ =1200 K is shown as black line.

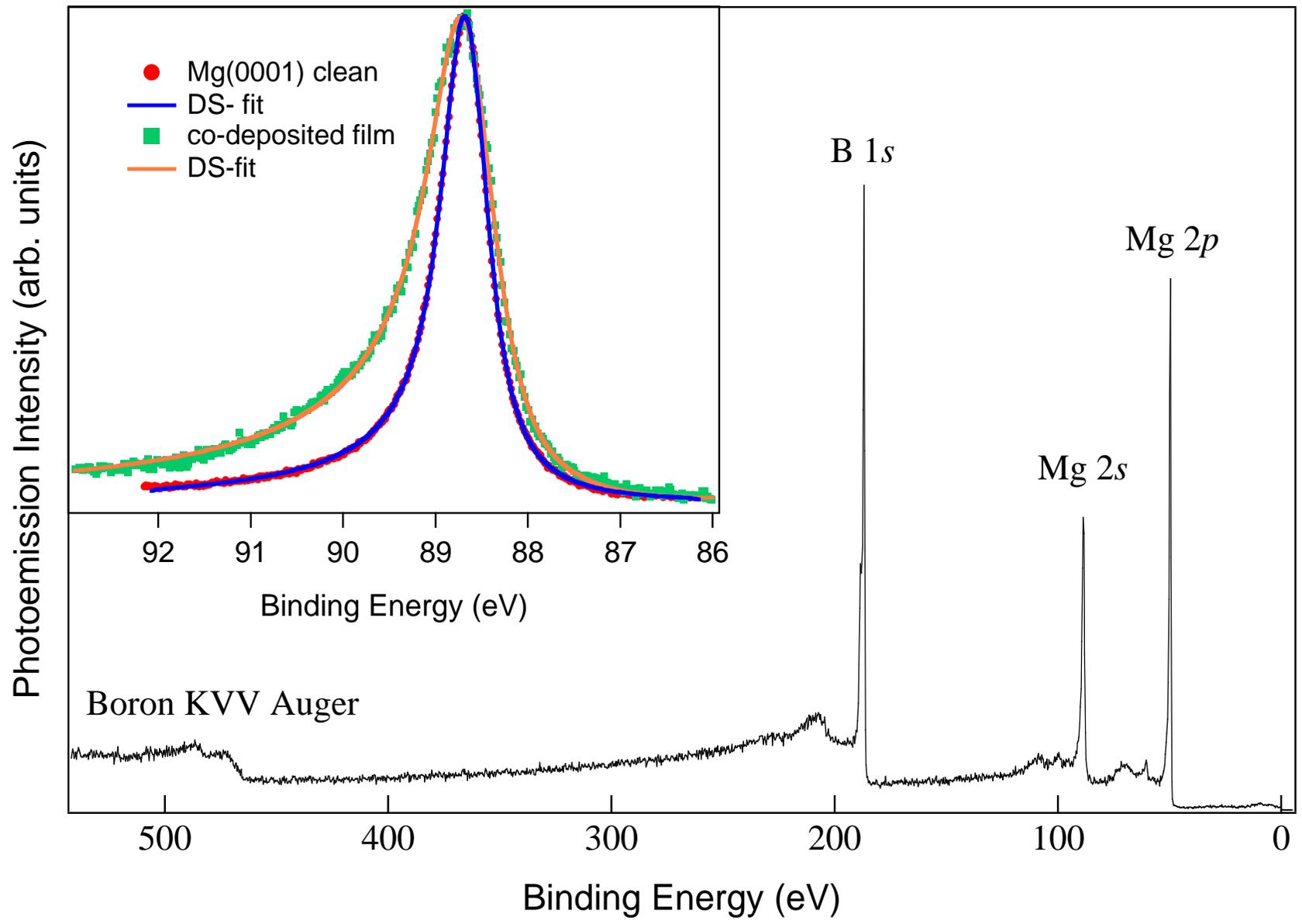

L. Petaccia et al. Fig. 1

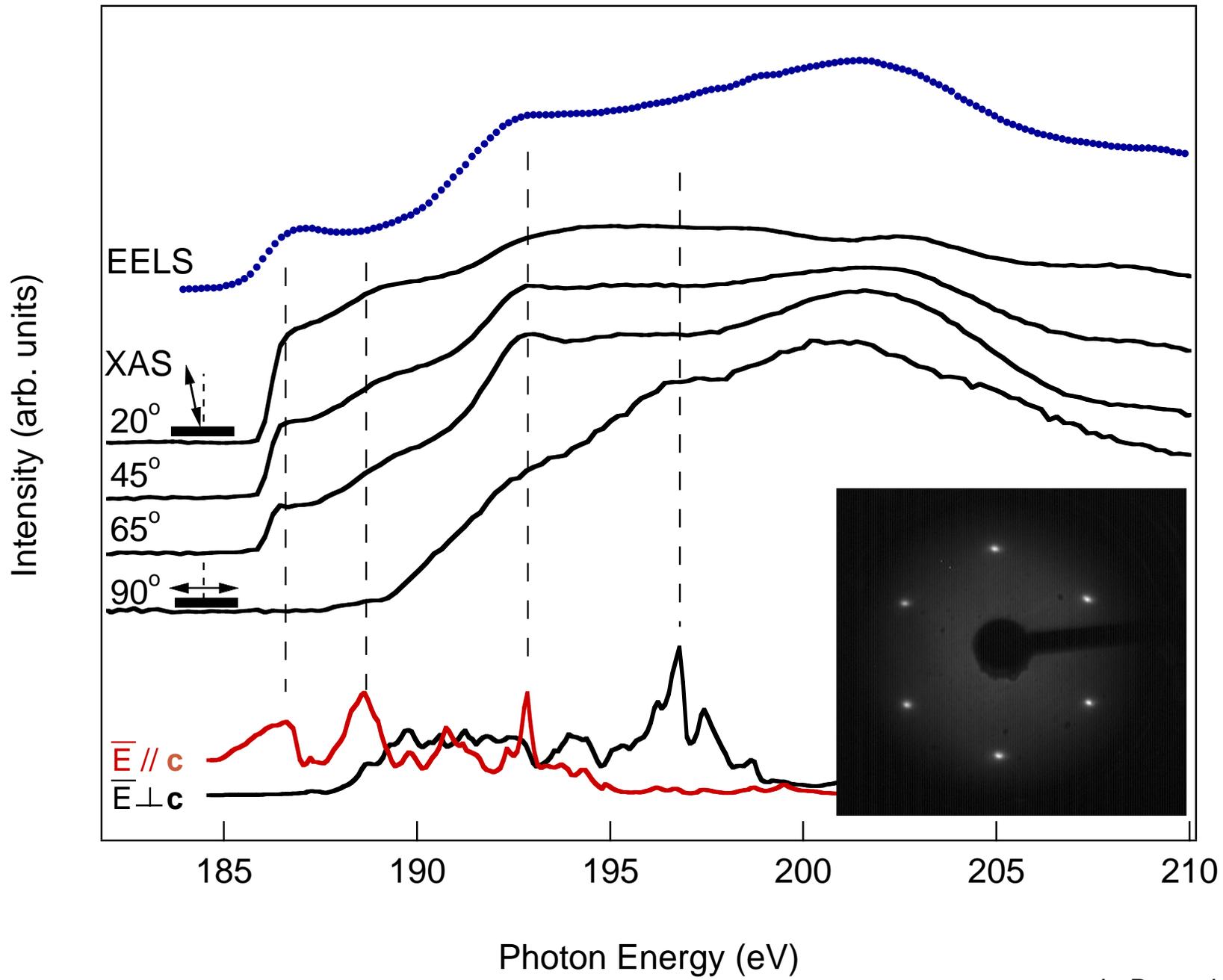

L. Petaccia et al., Fig. 2

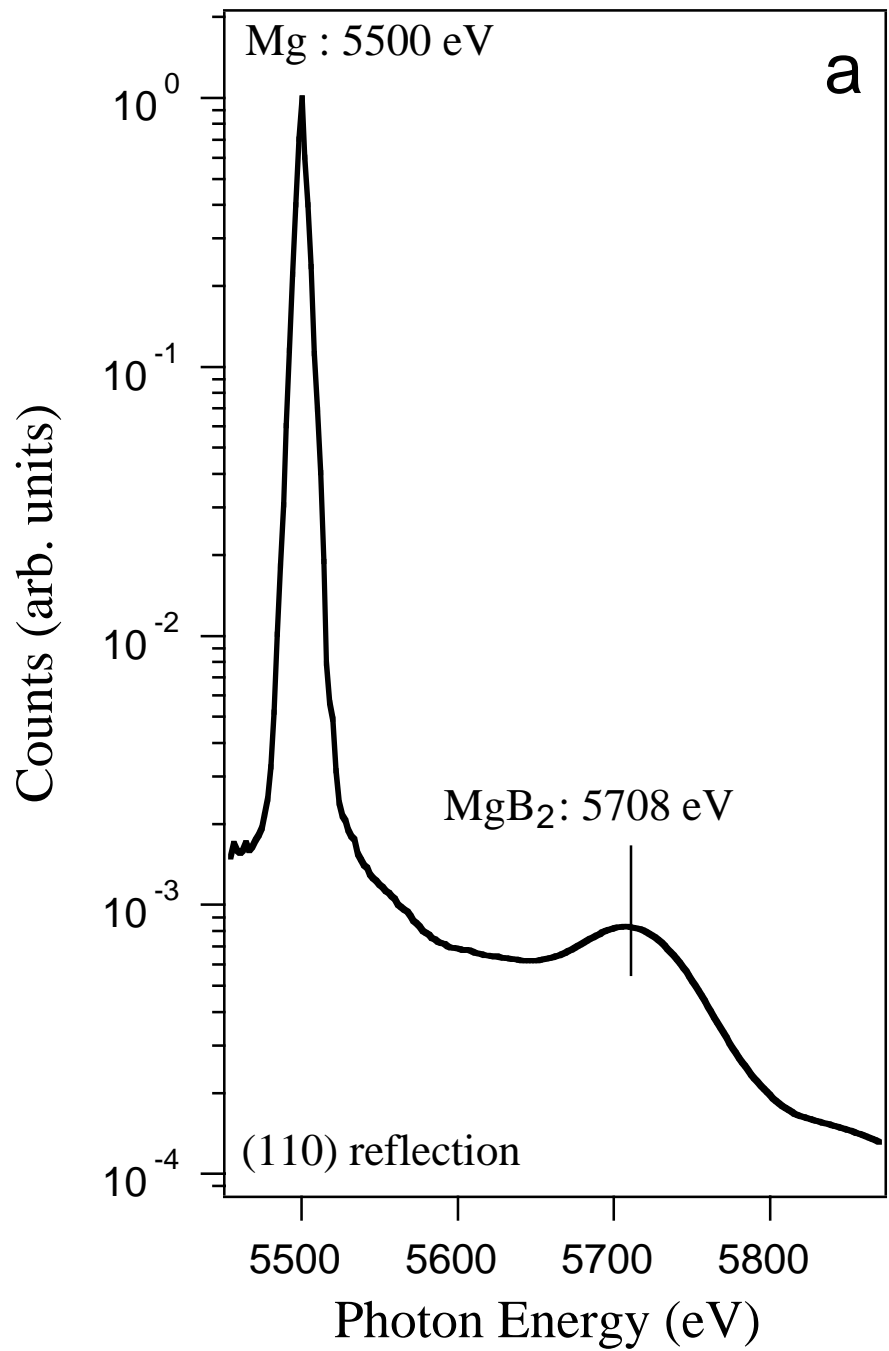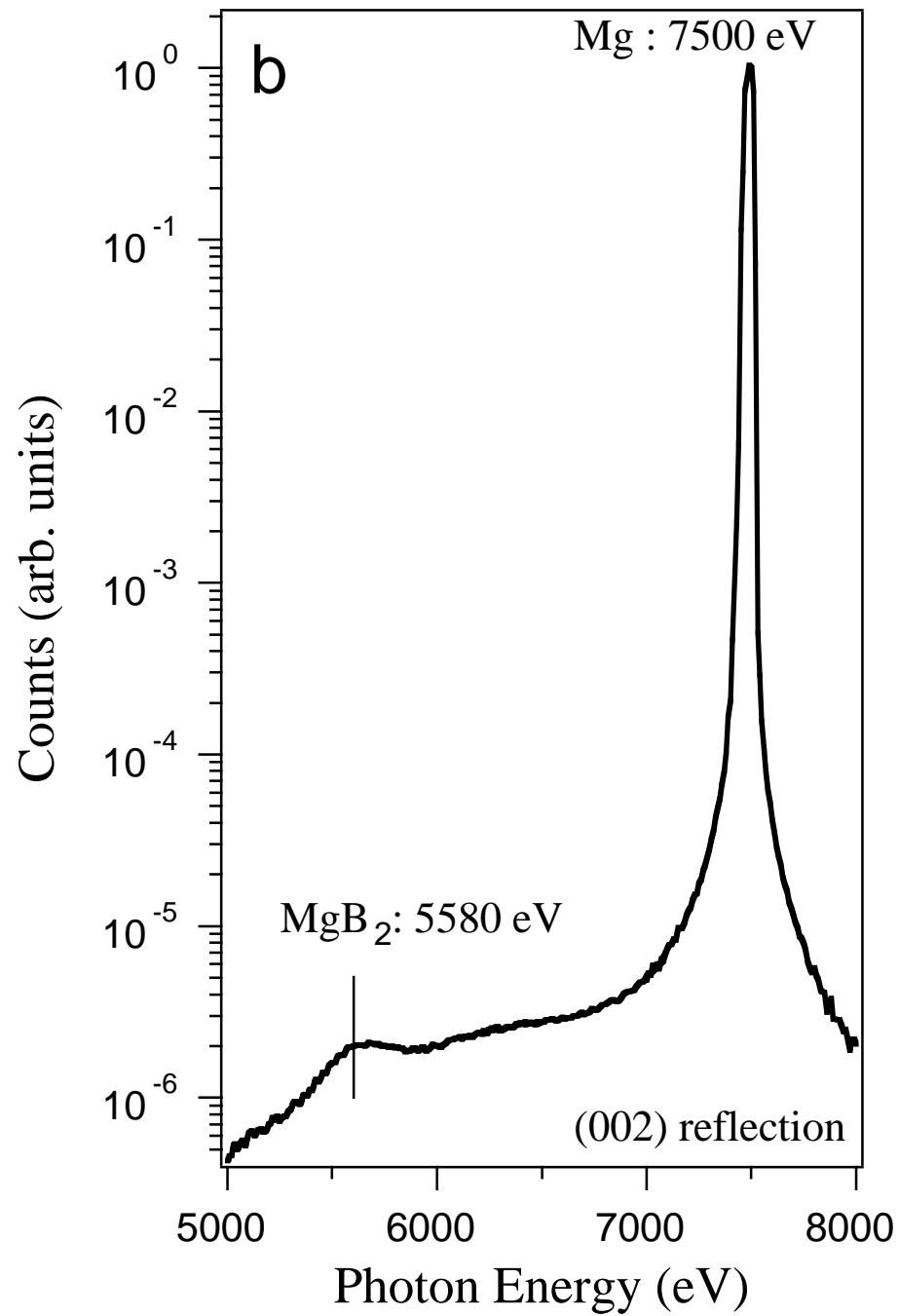

L. Petaccia et al.   Fig. 3

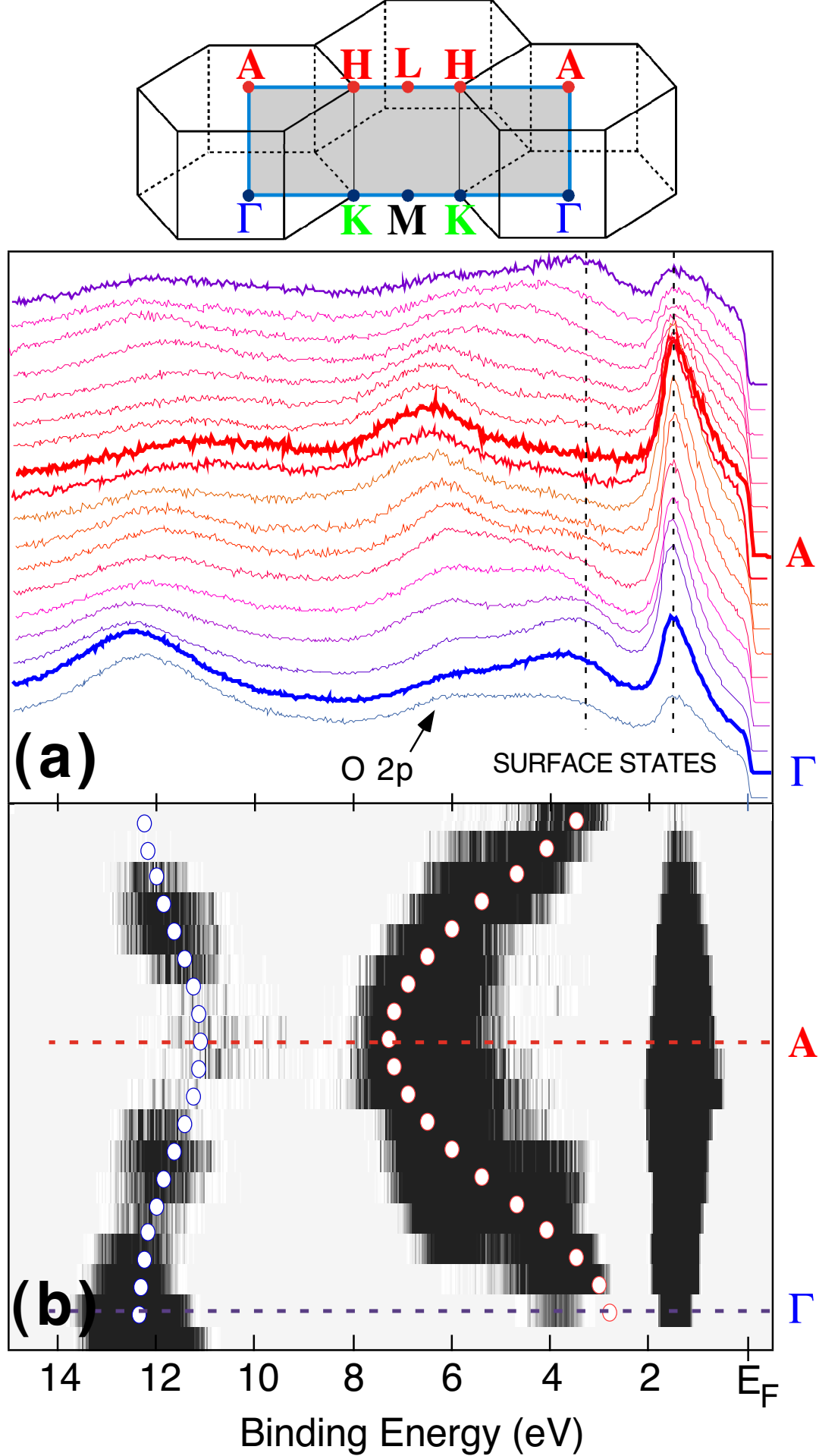

(a) O 2p SURFACE STATES

(b) Binding Energy (eV)

L. Petaccia et al., Fig. 4

(a)

Γ

K

M

K

(b)

Γ

12  8  4  $E_F$

Binding Energy (eV)

Photoemission Intensity (arb. units)

L. Petaccia et al.   Fig. 5

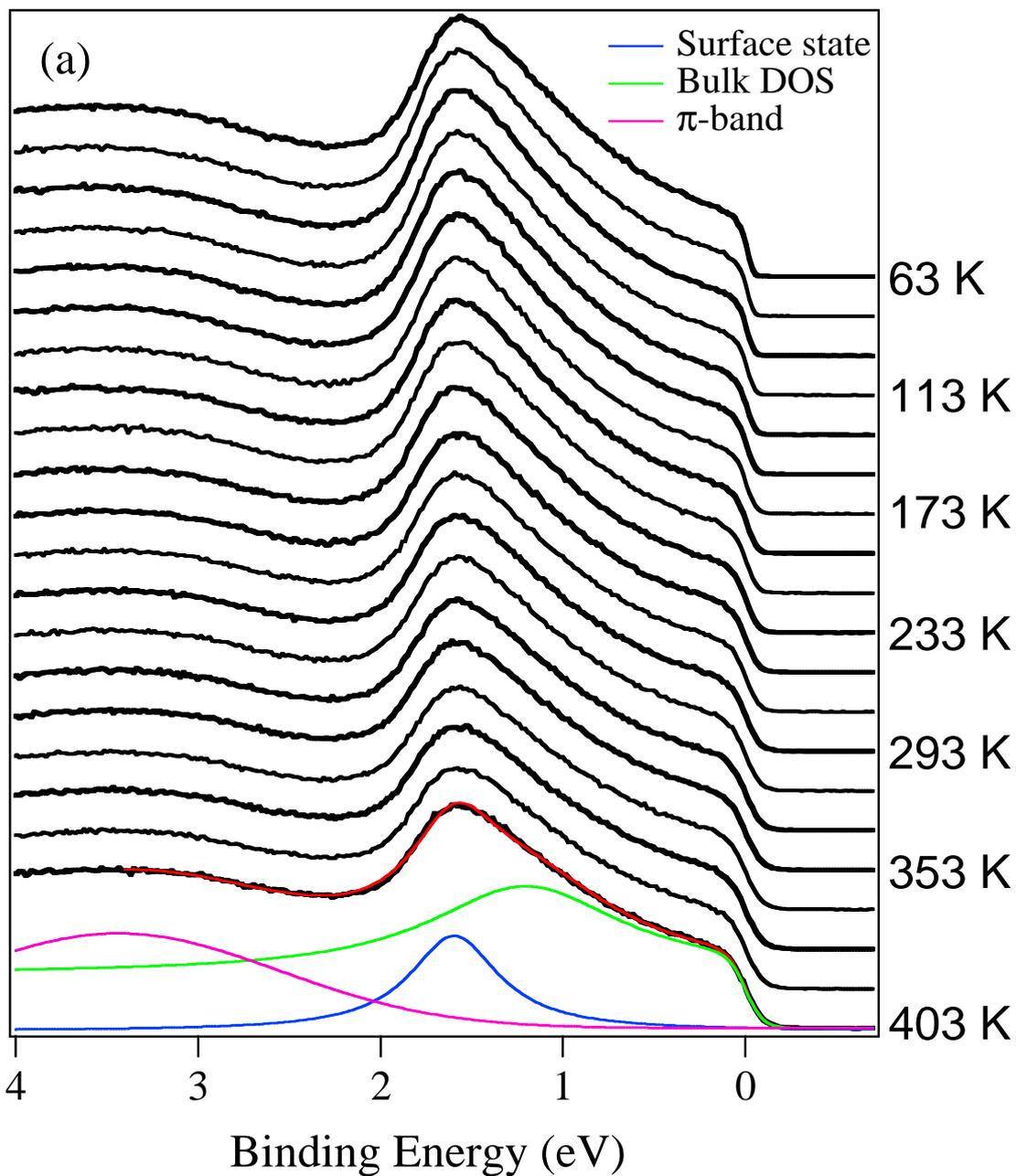

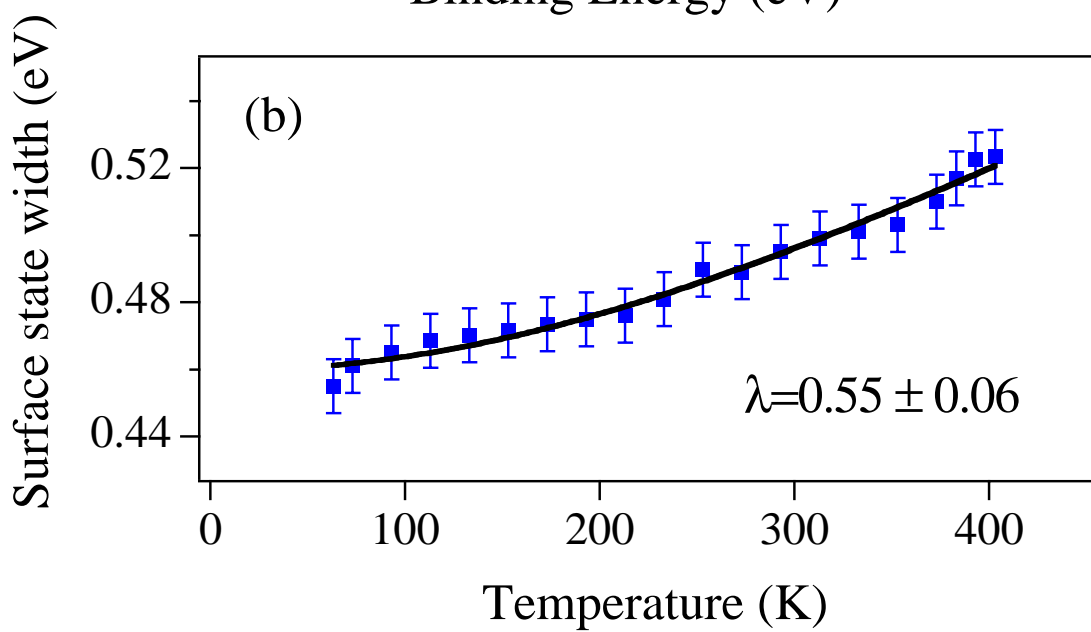

L. Petaccia et al.   Fig. 6